\def\arcsec{\hbox{$^{\prime\prime}$}}
\def\eps@scaling{.95}
\def\epsscale#1{\gdef\eps@scaling{#1}}

\def\plotone#1{\centering \leavevmode
    \epsfxsize=\eps@scaling\columnwidth \epsfbox{#1}}

\def\~{\thinspace}

\documentstyle[12pt,epsf,nature]{article}

\makeatletter
\def\thebibliography#1{\section*{References}
 \list
 {\arabic{enumi}.}{\settowidth\labelwidth{[#1]}\leftmargin\labelwidth
 \advance\leftmargin\labelsep
 \usecounter{enumi} \parsep 0.0ex}
 \def\newblock{\hskip .11em plus .33em minus -.07em}
 \sloppy\clubpenalty4000\widowpenalty4000
 \sfcode`\.=1000\relax}

\makeatother

\begin{document}

\title{\Large\bf Detection of Intergalactic Red Giant Branch Stars in the Virgo Cluster}
\author{ 
Henry C. Ferguson\\
Space Telescope Science Institute, \\
Baltimore, MD 21218, USA \\
ferguson@stsci.edu\\
\and
Nial R. Tanvir\\
University of Cambridge, Institute of Astronomy \\
Madingley Road, Cambridge CB3 0HA, UK \\
nrt@ast.cam.ac.uk\\
\and
Ted von Hippel\\
Department of Astronomy, University of Wisconsin \\
Address: WIYN Telescope, NOAO, PO Box 26732, \\
Tucson, AZ 85726, USA \\
ted@noao.edu
}

\maketitle

%\clearpage

{\bf
\baselineskip=12pt

It has been suspected for nearly 50 years that clusters of galaxies
contain a population of intergalactic stars, ripped from
galaxies during cluster formation or when the galaxies' orbits take
them through the cluster center.\cite{Zwicky51,Merritt84,MKLDO96} Support for the existence of such a population of
free-floating stars comes from measurements of the diffuse light in
clusters \cite{dV60,Gunn69,TK77,Uetal91,Vetal94,SK94}, and from recent detections of planetary nebulae with
positions and/or velocities far removed from any observed cluster
galaxy.  \cite{Aetal96,TW96} But estimates
for the mass of the diffuse population and its distribution relative to
the galaxies are still highly uncertain. Here we report the direct
detection of intergalactic stars in deep images of a blank field in the
Virgo Cluster. The data suggest that approximately
10\% of the stellar mass of the cluster is in intergalactic stars. 
We observe a relatively homogeneous distribution of stars, with evidence
of a slight gradient toward M87.
}

%\keywords{stars: and lots of them -- and far away, too}

%\section{Introduction}

%\baselineskip=24pt
\baselineskip=12pt
The process of cluster formation undoubtedly involved interactions and
mergers of galaxies and tidal ablation of stars due to the mean
gravitational shear of the cluster. \cite{Merritt84} Further
stripping is likely to have occured due to ``galaxy harassment'', the
cummulative effect of many close, high speed encounters with massive
cluster galaxies. \cite{MKLDO96} Theoretical predictions for the fraction of
mass in the intergalactic population range from 10\% to 70\%.  This
large uncertainty stems at a fundamental level from our poor knowledge
of when galaxies formed, when and how clusters collapsed, and how the
dark matter and baryonic matter were distributed within proto-galaxies.
\cite{Miller83,Merritt84,MKL96} Detection and mapping of
intergalactic stars could be an important tool for 
investigating these issues.

Searches for the diffuse light from intergalactic stars are extremely difficult due to the
stringent requirements on background subtraction.  Reported detections
and upper limits range from 10\% to 45\% of the total cluster light in
various clusters.  \cite{TK77,Uetal91,Vetal94,SK94}
Even when a signal is detected, it is unclear how much is due to a true
diffuse component, and how much to the extension of the galaxy
luminosity function below the limits of individual source detection.
Intergalactic planetary nebula candidates have been identified in the
Virgo cluster (3) and the Fornax cluster (10), but these numbers are
not yet sufficient to draw strong conclusions on the distribution of
intergalactic stars. 

%\section{Observations}

In May 1996 we used the Hubble Space Telescope WFPC2 camera with the
F814W (approximately I-band) filter to observe a blank field in the
Virgo Cluster (Fig. 1).  At faint limits compact galaxies are essentially
indistinguishable from stars, so the Hubble Deep Field, which was taken
in an area of sky with no foreground cluster, was used as a control.
For the comparisons, the HDF images were divided
into two groups, each consisting of exposures taken at three pointing
positions with total exposure times identical (to within 1\%) to the
Virgo Cluster field.  Because the sky noise was considerably lower in
the HDF images, additional noise was added to match the statistics in
the Virgo Cluster field.  These images were re-reduced and resampled to
match the Virgo Cluster images.

Figure 2 shows the comparison of source counts for the different
images.  Fainter than $I = 26.5$, there is a clear excess in source
counts in the Virgo Cluster image. The excess amounts to approximately
630 extra sources to $I=27.9$.  We expect fewer than 20 Galactic
foreground stars to this depth.  The major uncertainty is the
field-to-field variance in galaxy counts, because galaxies at the
limiting depth of our image are only marginally resolved. For a square
field of angular size $\theta$ on a side, the expected variance in the
number counts is $ \Delta N^2 = N + 2.24 N^2A\theta^{1-\gamma}$ where
$N$ is the mean number per field, and the angular correlation function
follows the powerlaw $w(\theta) = A \theta^{1-\gamma}$. \cite{Peebles75} Using N from the HDF, and adopting $A = 1.5 \times 10^{-4}$ and
$\gamma = 1.8$, \cite{BSM95} the expected standard deviation in
galaxy counts is $\Delta N = 80$, well below the 630 excess sources we
detect. As a check that the HDF is not highly unusual, source detection 
and aperture photometry were carried out on an additional
control field, centered on the radio galaxy 3C210. Source counts in
this field match the HDF to within 20\% down to the completeness
limit ($I = 27.5$) of the 3C210 field.

%\section{Discussion}

The total flux from the 630 excess sources detected in the 
image corresponds to a total magnitude
$I = 20.53$, or a surface brightness (dividing by the area of the
field) of $\mu_I = 31.16$ magnitudes per square arcsecond.
The source density is relatively uniform
across the field, with a marginally significant gradient towards M87.
To estimate the surface brightness and total mass of stars below
our detection limit,
consider a population with a
metallicity, expressed as a decimal logarithm relative to the solar iron-to-hydrogen
ratio, [Fe/H] = -0.7, an age of 13 Gyr, a Salpeter \cite{Salpeter55}  initial mass
function $(N(M) \propto M^{-2.35})$, and a distance 18.2 Mpc.
According to the theoretical luminosity function\cite{BBCFN94}
stars brighter than $M_I =  -3.4$ (corresponding to a limiting
magnitude $I = 27.9$ for the assumed distance) contribute 16\% of the
total flux from the stellar population. Our detection of 630 sources
thus implies an underlying surface brightness of $\mu_I = 29.14.$ The
theoretical color of the population is $B-I = 1.97$, so the $B$-band surface
brightness would be $\mu_B = 31.11;$  the surface mass density would be
$0.14 M_\odot \rm pc^{-2}$ if the initial mass function continues to $0.1 M_\odot$.

A purely empirical estimate of the surface brightness can be made by
comparison to previous HST observations of the galaxy
NGC\~3379.\cite{SMFLAB97} In those observations, approximately 39000
stars were detected in the three wide-field detectors, of which we estimate 5000
($\pm 50$\%) were brighter than $I = 26.9$ (roughly equivalent to our
detection limit in Virgo; see below). The NGC\~3379 surface brightness
at the radius of these observations is $\mu_B = 27.3$. \cite{CHLV90} The
difference in distance moduli between NGC3379 and the Virgo Cluster is
1.0 mag.\cite{TSFR95}
Thus our detection of 630 stars in Virgo implies a surface brightness
$-2.5 \log(630/5000) + 1.0$ magnitudes fainter, or $\mu_B = 30.6$, in
reasonable agreement with the theoretical determination, given the
uncertainties.

The intergalactic stars are likely to have originated primarily from the
elliptical and S0 galaxies in the cluster. The early-type galaxies
are more numerous in the cluster, have older stellar populations
(and hence higher stellar mass-to-light ratios),
and are likely to have inhabited the central megaparsec of the cluster
for much longer than the spirals and irregulars.\cite{TS84}  
To compare the intergalactic
light to the light emerging from the early-type galaxies, we divide the
Virgo cluster catalog \cite{BST85} into successively larger annuli 
of 70 galaxies centered on M87
and fit mean surface brightness vs. radius, where the mean surface
brightness is the total flux from galaxies divided by the area of each
annulus. This fit gives $\mu_B = 28.7$ at $r = 44.5^{\prime}$.
(This becomes $\mu_B = 27.9$ if spirals and irregulars are included.)

The implication is that about 10\% of the stellar mass in the cluster is in the 
intergalactic component. 

While the distinction between intracluster stars and the M87 halo
becomes purely semantical at a radius of 235 kpc, it is of some
interest to compare the inferred surface brightness to the extrapolated
surface brightness of M87.  An $r^{1/4}$-law fit to the Caon et al.
\cite{CCR90} photometry between 4.2 and 9.2 arcminutes, predicts a
surface brightness at $r=44.5$ arcminutes of $\mu_B = 30.3$ along the
major axis, and 33.4 along the minor axis. Our field is situated 25
degrees from the minor axis. Thus our detected number of stars is
higher than expected from a pure extrapolation of the visible portion
of M87. NGC\~4552,
located 28.8 arcmin from our field, could also contribute to the source counts. 
Even if we assume no tidal truncation, a
fit to the minor-axis photometry from 4.3 to 6.5 arcmin predicts $\mu_B
= 32.5$ at the position of our field, which would correspond to $\sim
25$\% of the detected source density.  Overlapping halos of galaxies
evidently account for only a portion of the stars we detect, unless
the outer profiles deviate substantially from the inner $r^{1/4}$
laws.

The estimated stellar mass fraction is sensitive to the age and
metallicity of the population. For a distance modulus $(M-m)_0 = 31.3$,
metallicity $[Fe/H] < -0.4$ and $\rm age > 1 Gyr$, the underlying
stellar mass fraction for our sample can plausibly range from 4 to
12\%.  At still younger ages the mass fraction goes down, while at
still higher metallicities it goes up.  Recent measurements of the
Virgo Cluster distance modulus range from $(m-M)_0 = 31.0$ to $31.7$.
\cite{FMMHFKSSGFHHHI94,TSFR95,ST96} The mass fraction is still
$\sim 5-10\%$ unless the distance modulus is at the high end of this
range {\it and} the metallicity is greater than $[Fe/H] = -0.7.$

The present observations thus favor a rather small mass fraction in the
diffuse population. In contrast, Theuns and Warren \cite{TW96} infer from the detection of 10 planetary nebula candidates, that
40\% of the stellar mass is in the diffuse population of the Fornax
Cluster.  \cite{TW96} Our results are consistent only if
the intergalactic population is metal-rich, or the Virgo Cluster is
more distant than $\sim 20$ Mpc. However, the planetary-nebula estimate
suffers from small number statistics, possible contamination of the
sample by background emission-line galaxies, and variations in the
planetary-nebula specific frequency with stellar population.

For an assumed distance of 18.2 Mpc, the total gravitating mass
estimated from X-ray observations in the central 240 kpc of the Virgo
Cluster is $1.2-3.5 \times 10^{13} M_\odot$, \cite{NB95} and the
corresponding mass to light ratio $M/L_B \approx 260$.  Most of the
{\it light} in the central region of the Virgo Cluster resides in the
central $10^{\prime}$ of M87, while most of the {\it mass} resides at
larger radii. For the region outside this central $10^{\prime}$, our
observations imply a mean $M/L_B \approx 700$.  The relatively smooth
distribution of mass inferred from the X-ray observations\cite{NB95} suggests that most of the intergalactic material was stripped via
tidal interactions with the cluster potential, although we cannot rule
out the possibility that some of the stars formed {\it in situ},
or that some were stripped off by impulsive interactions between galaxies
(``harassment'').

The luminosity of the tip of the red giant branch has been promoted as a distance indicator
of comparable precision to Cepheid variables. \cite{LFM93} While the
number of stars in our sample is not large enough for a precise
distance estimate, the shape of the observed luminosity function is
consistent with that seen in the halo of NGC\~5128 \cite{SMWGBBCCCGHHHSSTW96},
and in the halo of NGC\~3379 \cite{SMFLAB97}, shifted to a distance
modulus in the range $31.2 < (m-M)_0 < 31.6$.

Future observations should reveal whether the profile of the
intergalactic stars is similar to that of the gravitating mass, whether
it is centered on M87, and whether it is smoothly distributed.  The new
instruments on HST make it possible to measure the metallicity
distribution of the diffuse stellar population and compare it to the
metallicity distribution in the outer parts of elliptical galaxies and
in the X-ray gas. Star counts also open the possibility of measuring
the tidal radii of galaxies in the Virgo Cluster, and searching for
the wakes of galaxies moving through the intracluster material. The
combination of HST imaging and ground-based detection of planetary
nebulae provides a powerful new probe of the cluster environment and of
this nearly invisible population of stars.

\bibliography{naturemnemonic,bib}
\bibliographystyle{spie}

\section*{Methods}
The zero-point of the instrumental magnitude scale was established by
measuring the fluxes of the few bright isolated stars in the field in
$0.5 \arcsec$ radius apertures following the standard prescription.
\cite{Holtzman2} Additional corrections were made as follows:  (a)
$+0.05$ mag to correct for the so-called ``long exposure'' effect (Hill
et al. in press); (b) $-0.04$ mag, which is the appropriate color term
to convert to standard Cousin's $I$ magnitudes for red stars with
$V-I\sim1.5$;\cite{Holtzman2} and (c) $-0.04$ mag to allow for the
expected foreground extinction.\cite{BH84} We estimate that
the combined uncertainty from the zero-point calibration and the
various correction terms amounts to 0.06 mag.  Clearly extended sources
were removed from the catalog by restricting our analysis to sources
with a DAOPHOT sharpness parameter $-0.6 < s < 0.4$.  A similar
procedure was also followed for the HDF control field.  Finally, we
created 12 test data-sets in which 265 simulated stars were added to
the real data frames.  The input magnitudes of the simulated stars
covered the range from 23.5 to nearly 30.  We then processed these
frames in the same way as the original data and produced a matrix
relating the recovered magnitudes and detection efficiency to the input
magnitudes of the simulated stars.  From these simulations we estimate
that the catalog is $\sim$80\% complete at $I = 27.9$, and use this as
the limiting magnitude in our analysis.

\section*{Acknowledgments}

We are grateful to Shoko Sakai, Roberto Soria, and Carl Grillmair for
providing stellar photometry from nearby galaxies for comparison to the
Virgo Cluster data, to Mark Dickinson for the use of the 3C210
comparison field, and to Andy Fruchter and Richard Hook for use of the
drizzle code.  We would also like to thank Vera Rubin, Francois
Schweizer, and Abi Saha for helpful discussions and the Department of
Terrestrial Magnetism for hospitality (TvH).  TvH was partially
supported by a grant from the Edgar P.  and Nona B.  McKinney
Charitable Trust. Support for this work was also provided by NASA
through a General Observer research grant awarded by Space Telescope
Science Institute, which is operated by the Association of Universities
for Research in Astronomy.

\clearpage

\begin{figure}
\epsscale{0.7}
\plotone{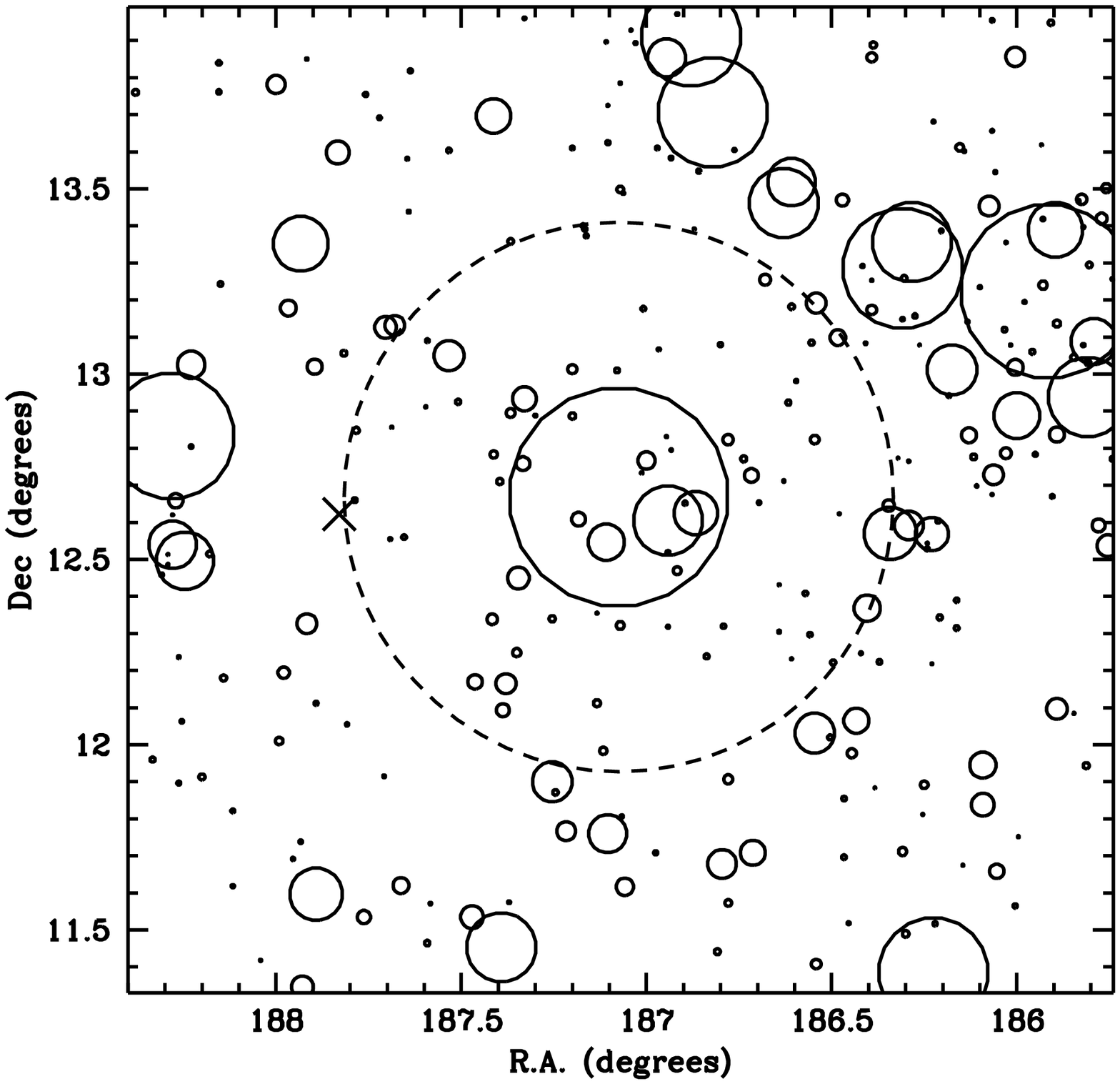}
\caption{\label{figfield}}
\end{figure}
\baselineskip=12pt
The 'X' marks the position of the HST images in this view of the core
of the Virgo Cluster. The solid circles mark positions of cluster
galaxies, and are scaled by the square root of the galaxy luminosity.
The central bright galaxy is M87, and the dashed circle marks a region
44.5 arcminutes from M87.  The field is located on this circle $\sim
235$ kpc east of M87 (J2000 coordinates $12^h 33^m 52^s, \,\,12^\circ
21^\prime 40^{\prime\prime}$) and well separated from any of the other
bright galaxies in the cluster (Figure \ref{figfield}).  The WFPC2
field covers 5.33 square arcminutes. The nearest
galaxy to the WFPC field is a dE galaxy with a magnitude B=18.0,
$6.6^\prime$ away. The nearest bright galaxy is NGC\~4552 $(B=10.8)$,
at a distance of $23^\prime$ ($=122$ kpc if the distance to the Virgo
Cluster is 18.2 Mpc).  The total exposure time for the observation was
33500 seconds. The exposures were taken at three positions separated by
$\pm 0.33 \arcsec$ to improve resolution and aid in removing hot pixels
and other cosmetic defects. Data were reduced and images combined
following the procedures for the Hubble Deep Field (HDF).
\cite{WBDDFFGGHKLLMPPAH96} The final reduced images were resampled via
``drizzling'' (Fruchter and Hook 1997, in preparation) onto a pixel
grid of $0.08 \arcsec$.

\clearpage
\begin{figure}
\epsscale{0.8}
\plotone{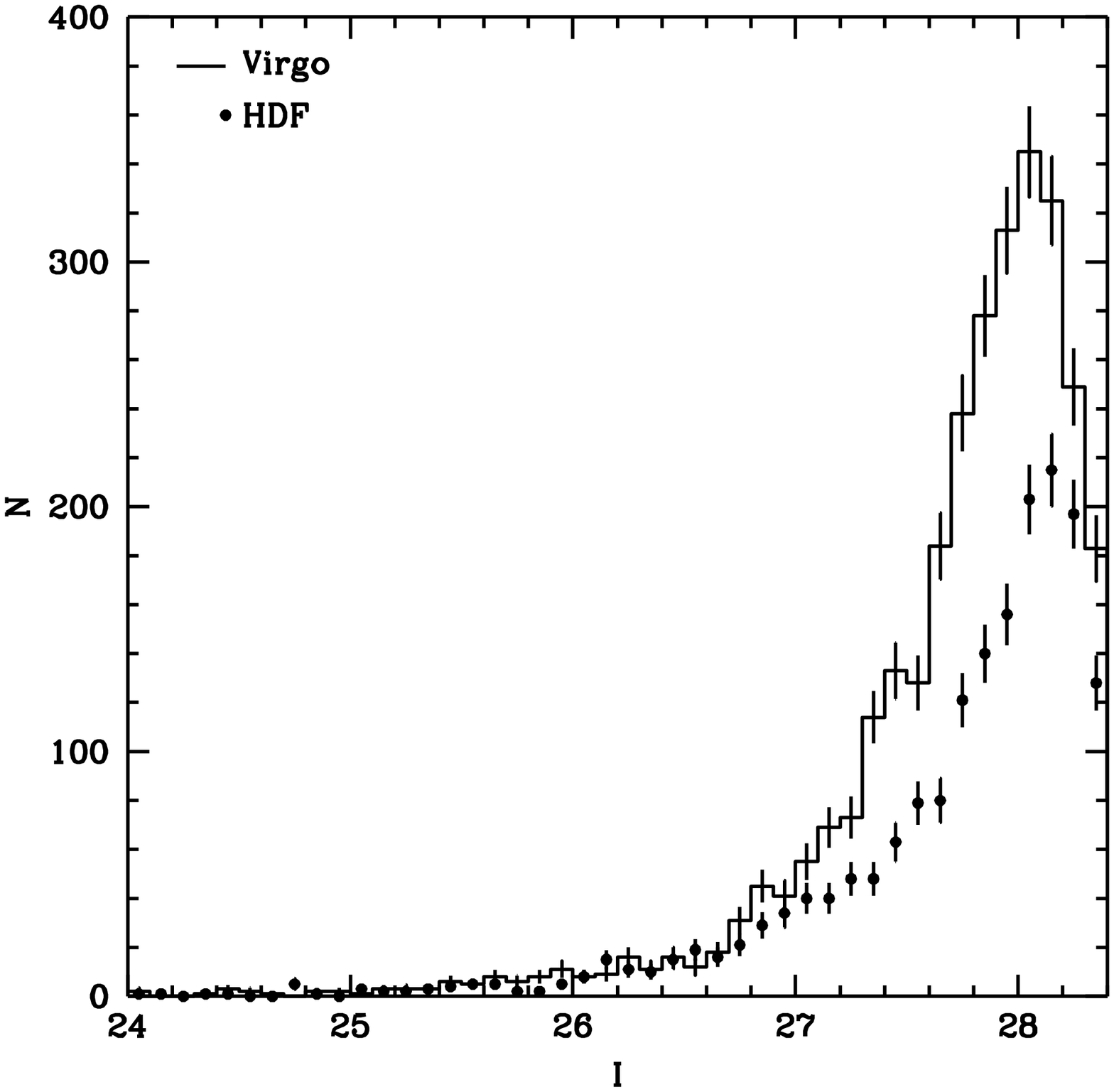}
\caption {\label{figcounts}}
\end{figure}
\baselineskip=12pt
Comparison of source counts in the Virgo Cluster field and the HDF.
The histogram is the Virgo Cluster field. The points with error bars
are the HDF. Magnitudes are based on PSF fitting. The excess in the
Virgo Cluster field begins at $I \approx 26.8$.  Source detection was
performed using DAOFIND, and profile-fitting photometry was performed
with the IRAF implementation of the DAOPHOT ALLSTAR code.  See Methods
for further details.

%\clearpage
%\plotone{xmarksthespot2.ps}
%\clearpage
%\plotone{virgohdf.ps}
\end{document}